\documentclass[12pt]{article}
\pdfoutput=1

\usepackage[utf8]{inputenc}
\usepackage{graphicx}
\usepackage{amsmath}
\usepackage{amssymb}
\usepackage{hyperref}
\hypersetup{colorlinks,citecolor=TokiwaIro,linkcolor=RuriIro,urlcolor=RuriIro,linktocpage}
\usepackage{cite}
\usepackage{braket}
\usepackage{bm}
\usepackage{bbm}
\usepackage{color}
\usepackage[svgnames,table]{xcolor}
\usepackage{float}
\usepackage{hhline}
\usepackage{multirow}
\usepackage{indentfirst}
\usepackage{latexsym}
\usepackage{mathtools}
\usepackage{cancel}
\usepackage{slashed}
\usepackage{colortbl}
\usepackage[mathscr]{euscript}
\usepackage{enumitem}
\usepackage{subcaption}
\usepackage{comment}
\usepackage{amsthm}

\usepackage{ulem}

\numberwithin{equation}{section}

\allowdisplaybreaks

\definecolor{RuriIro}{rgb}{0.,0.28,0.60}
\definecolor{TokiwaIro}{rgb}{0.,0.39,0.16}

\definecolor{kblue}{rgb}{0,0.48,0.73}
\definecolor{kred}{rgb}{0.73,0.25,0}
\definecolor{kgreen}{rgb}{0.48,0.73,0}

\theoremstyle{plain}

\theoremstyle{remark}

\newcommand{\nn}{\nonumber}

\newcommand{\pr}{{\prime}}

\newcommand{\eps}{\epsilon}
\newcommand{\Pcal}{\mathcal{P}}

\newcommand{\mc}{\mathcal}
\newcommand{\mr}{\mathrm}

\newcommand{\mbb}{\mathbb}

\newcommand{\ol}{\overline}
\newcommand{\dd}{\mathrm{d}}

\newcommand{\iu}{\mathrm{i}}

\newcommand{\kahler}{K\"{a}hler~}

\newcommand{\im}{\mathop{\mathrm{Im}}}
\newcommand{\TeV}{\mathop{\mathrm{TeV}}}
\newcommand{\GeV}{\mathop{\mathrm{GeV}}}
\newcommand{\MeV}{\mathop{\mathrm{MeV}}}

\newcommand{\FN}{{\mathrm{FN}}}

\newcommand{\MD}{\mathrm{MD}} 

\newcommand{\rta}{\mathop{\mathrm{Re}}\tau}
\newcommand{\ita}{\mathop{\mathrm{Im}}\tau}

\newcommand{\CW}{\mathrm{CW}}
\newcommand{\abs}[1]{\left|{#1}\right|}
\newcommand{\order}[1]{\mathcal{O} \left({#1}\right)}
\newcommand{\Wcal}{\mathcal{W}}
\newcommand{\eff}{\mathrm{eff}}
\newcommand{\Ykr}[2]{Y^{(#1)}_{#2}}
\newcommand{\la}{\lambda}

\begin{document}

\begin{titlepage}

\begin{flushright}
WU-HEP-26-02 \\
EPHOU-26-07
\end{flushright}

\vspace{1cm}

\begin{center}

{\LARGE \bfseries
  Finite modular Coleman--Weinberg inflation
}

\vspace{1cm}

\renewcommand{\thefootnote}{\fnsymbol{footnote}}
{%
\hypersetup{linkcolor=black}
Yoshihiko Abe$^{1,2,3}$\footnote[1]{yabe3@keio.jp},
\ 
Komei Goto$^{4}$\footnote[2]{komeigt@keio.jp},
\ 
Tetsutaro Higaki$^{4}$\footnote[3]{thigaki@rk.phys.keio.ac.jp},
\\
Junichiro Kawamura$^{5,6}$\footnote[4]{junichiro-k@ruri.waseda.jp},
and 
Tatsuo Kobayashi$^{7}$\footnote[5]{kobayashi@particle.sci.hokudai.ac.jp}
}%
\vspace{8mm}

{\itshape%
$^1${Graduate School of Science and Technology, Keio University, Yokohama, Kanagawa 223-8522, Japan.}\\
$^2${Keio University Sustainable Quantum Artificial Intelligence Center (KSQAIC), Keio University, Tokyo 108-8345, Japan.}\\
$^3${Quantum Computing Center, Keio University, 3-14-1 Hiyoshi, Kohoku-ku, Yokohama, Kanagawa, 223-8522, Japan.}\\
$^4${Department of Physics, Keio University, Yokohama 223-8533, Japan.}\\
$^5${Department of physics, Waseda University, Tokyo, 169-8555, Japan.}\\
$^6${Faculty of Health Science, Kumamoto Health Science University, Kumamoto 861-5598, Japan.}\\
$^7${Department of Physics, Hokkaido University, Sapporo 060-0810, Japan.}
}%

\vspace{8mm}

\end{center}

\abstract{
    We propose a modular symmetric inflationary model based on a Coleman--Weinberg potential generated by integrating out heavy vector-like quarks that couple to the complex modulus field $\tau$ through modular forms. 
    In this framework, the imaginary part of modulus $\tau$ plays the role of the inflaton, 
    while the real part is identified with a heavy axion. 
    We show that the model successfully explains the current cosmological observations.
    We further discuss reheating through modulus-dependent gauge kinetic functions and the cosmology of the axion. 
    The axion oscillation dominates over the Universe after the reheating via inflaton decay, 
    and then it decays before Big Bang Nucleosynthesis in the viable parameter region. 
    The quantum fluctuation of the axion can be of order $\mathcal{O}(1)\% $ of that of the inflaton,
    which would induce isocurvature perturbations that may be detectable in future observations. 
}

\end{titlepage}

\renewcommand{\thefootnote}{\arabic{footnote}}
\setcounter{footnote}{0}
\setcounter{page}{1}

\tableofcontents

\newpage 

\section{Introduction}

The origin of cosmic inflation in the early Universe remains 
one of the central open questions in particle physics and cosmology.
In the standard picture, inflation is driven by a scalar field, 
the inflaton, slowly rolling along a sufficiently flat potential, so that its vacuum energy dominates the expansion of the Universe.
The required flatness of the potential typically suggests 
the presence of an underlying symmetry 
that protects the inflaton mass from large quantum corrections.

In this work, we investigate a scenario 
in which the required flatness originates 
from the structure imposed by a finite modular symmetry $\Gamma_N$, 
with $N\in \mathbb{N}$ being called a level.   
Modular symmetry has been extensively studied to explain the flavor structure of quark and lepton masses in the Standard Model (SM)~\cite{Feruglio:2017spp}. 
Numerous flavor models based on finite modular symmetry successfully reproduce the observed masses and mixing angles~\cite{Kobayashi:2018vbk,Kobayashi:2018scp,Penedo:2018nmg,Novichkov:2018nkm,Feruglio:2021dte,Novichkov:2021evw,Petcov:2022fjf,Abe:2023ilq,Kikuchi:2023cap,Kikuchi:2023jap,Abe:2023qmr,Petcov:2023vws,deMedeirosVarzielas:2023crv,Abe:2023dvr}~\footnote{See also for reviews Refs.~\cite{Kobayashi:2023zzc,Ding:2023htn}.}.
Under modular symmetry and supersymmetry (SUSY), 
the Yukawa couplings are holomorphic functions 
of a complex modular field $\tau$, known as modular forms.
In string-inspired constructions, 
$\tau$ parametrizes the geometry of compact extra dimensions. 
Importantly, the modulus $\tau$ is a dynamical field 
whose vacuum expectation value (VEV) 
determines the Yukawa couplings and masses as well as other couplings.
This dynamical nature motivates us to explore the cosmological role of the modulus field.
As we discuss below, modular symmetry severely restricts the functional form of its potential.

The problem of moduli stabilization in modular-invariant frameworks has been extensively studied~\cite{Ferrara:1989bc,Font:1990nt,Cvetic:1991qm,Ferrara:1990ei}.
In models with finite modular symmetry $\Gamma_N$, stabilization mechanisms include flux-induced potentials~\cite{Ishiguro:2020tmo,Ishiguro:2022pde}, tree-level couplings to matter fields~\cite{Kobayashi:2019xvz,Kobayashi:2019uyt,Novichkov:2022wvg,Knapp-Perez:2023nty,Kobayashi:2023spx,Abe:2024tox,Higaki:2024pql}, and radiative corrections~\cite{Higaki:2024jdk,Funakoshi:2024yxg}.
The first two mechanisms typically stabilize the modulus 
at the fixed points $\tau = \iu$ and $\tau = \omega := e^{2\pi \iu/3}$,
while stabilization in the asymptotic region $\tau\to \iu\infty$ can be realized 
for the radiative corrections of heavy modes whose masses are determined 
by the non-trivial singlet modular forms~\cite{Higaki:2024jdk}.
The inflation scenarios on the moduli potential 
has been studied, 
although the discussion focused on the modulus trajectories 
near the fixed points $\tau = i$ or $\tau=w$~\cite{Kobayashi:2016mzg,Higaki:2015kta,Schimmrigk:2016bde,Schimmrigk:2021tlv,Abe:2023ylh,Casas:2024jbw,Aoki:2025wld,Ding:2024neh,King:2024ssx,Ding:2024euc,Jiang:2025qbi}.

The region near $\tau = \iu\infty$ is particularly intriguing.
It naturally explains flavor hierarchies via a Froggatt–Nielsen (FN) mechanism~\cite{Abe:2023ilq,Abe:2023qmr,Abe:2023dvr,Kikuchi:2023fpl}.
In this asymptotic limit, modular forms admit a $q$-expansion with $q = e^{2\pi \iu \tau}$, 
which results in an exponentially suppressed slope of the scalar potential~\footnote{
The double-exponent type inflaton potential is studied in the $\alpha$ attractor model~\cite{Kallosh:2024kgt,Kallosh:2024ymt}.
}.
After canonical normalization, 
$\phi \sim \ln(\ita)$, the potential takes a double-exponential form, 
$V \sim \mathrm{exp}(-C e^\phi)$, with a positive constant $C$.
This structure leads to small slow-roll parameters, 
consistent with current cosmological constraints.
Therefore, the modulus dynamics in the large $\ita$ regime 
provide a natural realization of slow-roll inflation~\cite{Abe:2023ylh}.
In string compactifications, 
this regime corresponds to the large-volume limit, 
where the rescaling symmetry $\tau \to \lambda \tau$, $\lambda\in\mathbb{R}$, emerges 
as an approximate symmetry. 
This gives rise to an emergent shift symmetry of the canonically normalized inflaton, 
$\phi \sim \ln \ita \to \phi + \ln \lambda$, 
which protects the flatness of the inflaton potential~\cite{Cicoli:2023opf,Cicoli:2026bqo}. 
The double-exponential potential can then be interpreted as a radiatively induced breaking of this emergent shift symmetry.

In this article, we investigate slow-roll inflation driven by the radiatively generated modulus potential.
We demonstrate that our model successfully reproduces the observed primordial power spectrum due to the exponentially suppressed slope of the potential~\cite{Planck:2018jri,Planck:2018vyg}.
We also show that the reheating temperature is sufficiently high to be consistent with Big Bang Nucleosynthesis (BBN).
Furthermore, we discuss the phenomenology of the $\rta$ direction, which behaves as an axion-like field~\cite{Higaki:2024jdk,Jung:2024bgi}. 
The axion mass is much lighter than the Hubble scale
due to the approximate shift symmetry $\tau \to \tau + c$ which is originated 
from the residual discrete symmetry $\tau \to \tau + 1$ of the finite modular symmetry $\Gamma_N$.

The rest of this paper is organized as follows.
We briefly review finite modular symmetry in Sec.~\ref{sec:modular-sym}.
Our inflation model based on the radiative modulus potential is introduced in Sec.~\ref{sec:inflation-model}, where we also present numerical results and discuss its consistency with observations.
In Sec.~\ref{sec:reheating}, we discuss reheating after inflation.
The axion phenomenology is examined in Sec.~\ref{sec:axion-cos}.
Sec.~\ref{sec:conclusion} is devoted to our conclusions.
The explicit form of the derivatives of the normalized potential is shown in App.~\ref{app:derivatives},
and the axion abundance produced from the inflaton decay is discussed in App.~\ref{app:axion-abundance}.

\section{Finite modular symmetry}
\label{sec:modular-sym}

A finite modular group $\Gamma_N$ is defined as the quotient group 
of the special linear group  
$\bar{\Gamma}\coloneqq\mathrm{PSL}(2,\mathbb{Z})=\mathrm{SL}(2,\mathbb{Z})/\{\pm1\}$
divided by the principal congruence group $\Gamma(N)$:
\begin{equation}
    \Gamma_N \coloneqq \bar{\Gamma}/\Gamma(N),
\end{equation}
where $\Gamma(N)$ is defined as 
\begin{equation}
    \Gamma(N) \coloneqq \left\{
        \left.
            \begin{pmatrix}
                a & b \\ c & d
            \end{pmatrix}
            \in \overline{\Gamma}\ 
        \right|
        \begin{pmatrix}
            a & b \\ c & d
        \end{pmatrix}
        \equiv
        \begin{pmatrix}
            1 & 0 \\ 0 & 1
        \end{pmatrix} \quad \mr{mod}\,N
    \right\}.
\end{equation}
Here, $N\in\mbb{N}$ is the level.  
The modular transformation 
$\gamma \in SL(2,\mathbb{Z})$
for a complex parameter $\tau$, with $\im\tau>0$, is given by
\begin{equation}
    \tau \to \gamma \tau =\frac{a\tau+b}{c\tau +d}.
\end{equation}
This transformation is generated by
\begin{equation}
    S=\begin{pmatrix}
        0 & 1 \\ -1 & 0
    \end{pmatrix},
    \quad
    T=\begin{pmatrix}
        1 & 1 \\ 0 & 1
    \end{pmatrix},
\end{equation}
which satisfy the algebraic relations
\begin{equation}
    S^2 = (ST)^3 = T^N = 1,
\end{equation}
for $\Gamma_N$.
This algebra illustrates 
that finite modular symmetries are isomorphic to the non-Abelian discrete symmetries,
$\Gamma_{2n} \sim S_{n+2}$, $\Gamma_{2n+1} \sim A_{n+3}$ for $n=1,2$. 
There are discrete Abelian symmetries, 
$\mbb{Z}_2^S,\,\mbb{Z}_3^{ST}$, and $\mbb{Z}_N^T$ in the algebra, 
whose fixed points are $\tau =\iu$, $\omega := e^{2\pi \iu/3}$, and $\iu\infty$, respectively.
In other words, the corresponding residual symmetry remains unbroken 
when the modulus is stabilized at one of these fixed points.

A modular form $Y_r^{(k)}(\tau)$ 
with a modular weight $k$ and representation $r$ 
is a holomorphic function of $\tau$ transformed as
\begin{equation}
    Y_r^{(k)}(\tau) \rightarrow 
    Y_r^{(k)}(\gamma\tau) = (c\tau+d)^k \rho_r(\gamma) Y_r^{(k)}(\tau),
\end{equation}
where 
$\rho_r(\gamma)$ is a unitary matrix of a representation $r$.  
For example, the triplet representation of weight $2$ 
under $\Gamma_3 \simeq A_4$ is given by~\cite{Feruglio:2017spp,Kobayashi:2023zzc,Kobayashi:2023spx,Higaki:2024jdk} 
\begin{align}
    Y_3^{(2)}(\tau) = (Y_1(\tau), Y_2(\tau), Y_3(\tau)),
\end{align}
where 
\begin{align}
    Y_1(\tau) &= \frac{\iu}{2\pi} \biggl[
        \frac{\eta'(\tau/3)}{\eta(\tau/3)} + \frac{\eta'((\tau+1)/3)}{\eta((\tau+1)/3)} + \frac{\eta'((\tau+2)/3)}{\eta((\tau+2)/3)} - \frac{27 \eta'(3\tau)}{\eta(3\tau)}
    \biggr],
    \\
    Y_2(\tau) &= \frac{-\iu}{\pi} \biggl[
        \frac{\eta'(\tau/3)}{\eta(\tau/3)} 
        + \omega^2 \frac{\eta'((\tau+1)/3)}{\eta((\tau+1)/3)} 
        + \omega \frac{\eta'((\tau+2)/3)}{\eta((\tau+2)/3)}
    \biggr],
    \\
    Y_3(\tau) &= \frac{-\iu}{\pi} \biggl[
        \frac{\eta'(\tau/3)}{\eta(\tau/3)} 
        + \omega \frac{\eta'((\tau+1)/3)}{\eta((\tau+1)/3)} 
        + \omega^2 \frac{\eta'((\tau+2)/3)}{\eta((\tau+2)/3)}
    \biggr]. 
\end{align}
Here, $\eta(\tau)$ is the Dedekind $\eta$-function defined as
\begin{align}
    \eta(\tau) := q^{1/24} \prod_{n=1}^\infty (1-q^n),
    \qquad 
    q \coloneqq e^{2\pi \iu \tau}.
    \label{eq:Dedekind-eta}
\end{align}
We can construct the modular forms with higher weights. 
In our analysis, we use the non-trivial singlets given by 
\begin{align}
\label{eq:def_Y12}
    Y_{1_1}^{(4)}(\tau) &= -\frac{1}{12} \left(Y_3^2 + 2 Y_1 Y_2\right), 
    \\ \notag 
    Y_{1_1}^{(12)}(\tau) &= -\frac{1}{12} (Y_1^2 + 2 Y_2 Y_3)^2 (Y_3^2 + 2 Y_1 Y_2), 
    \\ \notag 
    Y_{1_2}^{(12)}(\tau) &= \frac{1}{144}(Y_1^2 + 2 Y_2 Y_3) (Y_3^2 + 2 Y_1 Y_2)^2, 
\end{align}
where the non-trivial singlet $1_t$ 
transforms as $1_t \to w^t 1_t$ under the $T$ transformation $\tau \to \tau + 1$, 
whereas it is not transformed by the $S$. 
The $q$-expansion of the modular forms is given by
\begin{align} 
\label{eq-ExYsinglets}
    Y_{1_1}^{(4)}(\tau) &= 
    q^{1/3} \left( 1 - 8 q + 20 q^2 + \mathcal{O}(q^3) \right), 
    \\ \notag 
    Y_{1_1}^{(12)}(\tau) &= q^{1/3}\left( 1 + 472 q +551040 q^2 + \mathcal{O}(q^3)
      \right), 
    \\ \notag 
    Y_{1_2}^{(12)}(\tau) &= q^{2/3} \left(1 + 224 q - 1576 q^2 + \mathcal{O}(q^3)
     \right), 
\end{align}
which become accurate for larger $\ita$ since $\abs{q} = e^{-2\pi \ita}$. 
We choose the normalization of the modular forms 
in Eq.~\eqref{eq:def_Y12} 
such that the coefficients of the first terms in the $q$-expansion are unity.
Note that the power of the leading term 
is determined by the $T$-charge $t$ under the $\mathbb{Z}^T_N$ symmetry.

We assume that a chiral superfield $Q$ with modular weight $-k_Q$ and representation $r_Q$ transforms as~\cite{Feruglio:2017spp,Kobayashi:2023zzc} 
\begin{align}
    Q \mapsto (c\tau+d)^{-k_Q} \rho(r_Q) Q, 
\end{align}
so that the interactions are invariant under the modular transformation.

\section{Finite modular CW inflation}
\label{sec:inflation-model}

We consider a supersymmetric model with the following \kahler potential 
and superpotential~\cite{Kobayashi:2023spx,Higaki:2024jdk,Higaki:2024ueb}:
\begin{align} 
\label{eq:model_Lagrangian}
        K &= -h\log{(-\iu\tau+\iu\tau^\dag)}
            +\sum_i 
            \left(
                \frac{Q^\dagger_i Q_i}{(-\iu\tau+\iu\tau^\dag)^{k_{Q_i}}}
                +
                \frac{\bar{Q}^\dagger_i \bar{Q}_i}{(-\iu\tau+\iu\tau^\dag)^{k_{\bar{Q}_i}}}
            \right),
     \notag \\
        W &= \sum_i M_{Q_i} Y_{r_i}^{(k_i)}(\tau) \bar{Q}_i Q_i,
\end{align}
where $h$ is a natural number and we use the unit $M_p=1$, where $M_p$ denotes the reduced Planck scale.
For simplicity, we assume that $Q_i$ and $\bar Q_i$ transform as singlet representations of $\Gamma_N$ or its double cover $\Gamma_N'$ associated with $SL(2,\mathbb Z)$.
The chiral superfields $Q_i$ and $\bar{Q}_i$ are pairs of vector-like quarks (VLQs) under the SM gauge group, 
with the modular weights $-k_{Q_i}$ and $-k_{\bar{Q}_i}$, respectively. 
Note that we adopt the simplest form of the K\"ahler potential in Eq.~\eqref{eq:model_Lagrangian}, 
which may be realized in the large-volume limit of string compactifications. 
The K\"ahler potential exhibits the rescaling symmetry $\tau \to \la \tau$, 
together with $\Phi \to \la^{k_\Phi/2} \Phi$ ($\Phi = Q_i, \ol{Q}_i$), where $0<\la\in\mathbb{R}$, 
up to a K\"ahler transformation~\footnote{
Equivalently, the full model is invariant if one assigns the transformation $W\to\lambda^{h/2}W$.
At the level of the kinetic terms derived from the leading K\"ahler potential, the rescaling symmetry, together with the shift symmetry $\tau \to \tau + c$, may be viewed as part of a continuous $SL(2,\mathbb{R})$ symmetry, of which the modular symmetry $SL(2,\mathbb{Z})$ is a discrete subgroup.}. 
This rescaling symmetry can be read as the shift symmetry of the canonically normalized inflaton,  
$\phi \sim \log \ita \to \phi + \log \la $ as shown in Sec.~\ref{sec-slowroll}. 
This emergent shift symmetry of the inflaton is broken by the superpotential 
which generates the dominant inflaton potential.

Their masses are characterized by the parameter $M_{Q_i}$, 
whose natural scale may be around the Planck scale $M_p$.
The modular weight of $Y^{(k_i)}_{r_i}$ satisfies 
$k_i = k_{Q_i} + k_{\bar{Q}_i} - h$, 
so that the supergravity action is invariant 
under the modular transformation~\cite{Ferrara:1989bc}.  
The index $r_i$ is the representation of the modular form $Y^{(k_i)}_{r_i} (\tau)$ which transforms under the modular symmetry $\Gamma_N$~\footnote{ 
When $h$ is an odd number, at least one of fields should have an odd weight to satisfy condition. 
In such a case, we have to consider the double covering $\Gamma_N^\pr$ to realize modular forms with odd weights.
}. 
For illustration, we consider the three non-trivial singlet representations of $\Gamma_{3}$, 
namely $Y^{(4)}_{1_1}$, $Y^{(12)}_{1_1}$, and $Y^{(12)}_{1_2}$. 
Here, $\Ykr{4}{1_1}$ is the non-trivial singlet with the smallest weight, 
and we also study the modular weights $k=12$, especially $\Ykr{12}{1_2}$ in our cosmological study, 
to see the effects of the large weight and different representation. 
To have the modular form with $r_i = 1_t$, 
we assign the representation of the matter fields as $r_Q = r_{\ol{Q}} = 1_t$,
so that $1_t \otimes 1_t \otimes 1_t = 1_0$ under $\Gamma_3$ and the mass term in the superpotential is allowed. 
We will study the potential of the modulus $\tau$ 
generated by the loop corrections in the following.

\subsection{CW potential and its minimum}

After integrating out the heavy modes $Q_i$ and $\bar{Q}_i$,
the vector-like mass term in Eq.~\eqref{eq:model_Lagrangian} generates the 1-loop Coleman-Weinberg (CW) 
potential~\cite{Higaki:2024jdk}, 
\begin{equation}\label{eq:Coleman-Weinberg}
    \begin{split}
        V_{\mr{CW}}
        = \frac{1}{32\pi^2}\sum_{i}
        \Bigg[
            \left(
                m_i^2 + m_{Q_i}^2(\tau)
            \right)^2
            &
            \left(
                \log{\frac{m_i^2 + m_{Q_i}^2(\tau)}{\mu^2}}
                -
                \frac{3}{2}
            \right)
            \\
            &-
            (m_{Q_i}^2(\tau))^2
            \left(
                \log{\frac{m_{Q_i}^2(\tau)}{\mu^2}}
                -
                \frac{3}{2}
            \right)
        \Bigg],
    \end{split}
\end{equation}
where $m_i$ is the soft supersymmetry breaking mass of $Q_i$, assumed to be independent of the modulus $\tau$.
Here, $m_{Q_i}^2$ corresponds to the physical mass squared 
of the vector-like quark $Q_i$, 
and is a function of $\tau$ whose form is given by 
\begin{equation}
\label{eq:m_Qi}
    m_{Q_i}^2(\tau) 
    = M_{Q_i}^2(-i\tau+i\tau^{\dagger})^{k_{i}}
    \left| Y^{(k_i)}_{r_i}(\tau) \right|^2.
\end{equation}
Note that the physical mass $m_{Q_i}^2(\tau)$ after the canonical normalization is modular invariant.
The parameter $\mu$ denotes the renormalization scale and
characterizes the physical mass of the VLQ after the stabilization.
Throughout this paper, 
we assume that the mass parameters are universal for $Q_i$'s 
and hence we omit the indices, 
$m_{Q_i}(\tau) = m_Q(\tau)$, $M_{Q_i} = M_Q$, $m_i = m_0$, and $k_i = k$ 
in the following.

\begin{figure}[t]
   \centering
   \includegraphics[width=0.65\linewidth]{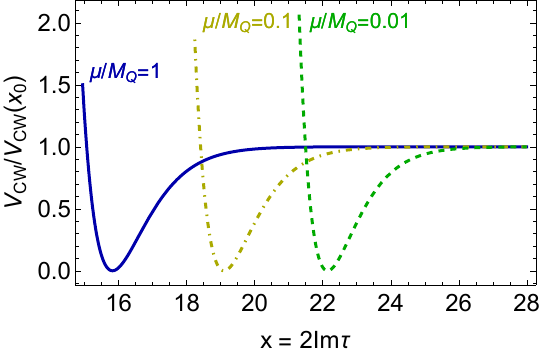}
   \caption{
        We show the normalized CW potential 
        with $t=1$, $k=12$, $N=3$ and $m_0/M_Q = 10^{-5}$ 
        The blue solid, yellow dot-dashed and green dashed lines are $\mu/M_Q = 1$, $10^{-1}$ and $10^{-2}$, respectively.
   }
   \label{f_CWpotential}
\end{figure}

The CW potential Eq.~\eqref{eq:Coleman-Weinberg}
can be read as the potential for the modulus $\tau$. 
As pointed out in Refs.~\cite{Higaki:2024jdk,Higaki:2024ueb},
the potential has a minimum at $\im\tau > 1$ if $m_0^2 \ll m_Q^2$. 
To see this, we expand the potential with respect to $q:= e^{2\pi\iu \tau}$, which is small for $\im\tau \gg 1$. 
In general, a singlet modular form of $\Gamma_N$ is expanded as
\begin{equation}
\label{eq:qexpansion}
    Y^{(k)}_{1_t} = q^{{t}/{N}} \sum_{n=0}^\infty c_n q^n, 
\end{equation}
where we employ the normalization of the modular forms 
so that $c_0 = 1$ as in Eq.~\eqref{eq:def_Y12}. 
Here, $t$ denotes the charge of the singlet modular form 
under $\mbb{Z}_N^T$ symmetry and is $0\leq t \leq N-1$. 
The explicit forms of the singlets under $\Gamma_3$ are 
shown in Eq.~\eqref{eq-ExYsinglets}.
By inserting this into Eq.~\eqref{eq:Coleman-Weinberg}, 
the CW potential at the leading order in $q$ and $m_0$ is given by 
\begin{equation} \label{eq:Vcw_infty}
    V_{\mr{CW}}(x)  
    = \frac{m_0^2 M_Q^2}{16\pi^2} 
        \left[ x^k e^{-px}
        \left(
            \log{\frac{M_Q^2}{\mu^2}+k\log{x}-px-1}
        \right)
        +\order{|q|,\,\frac{m_0^2}{M_Q^2}}  
        \right],
\end{equation}
where $x\coloneq2\im{\tau}$ and $p\coloneqq 2\pi t /N$.
Its first derivative with respect to $x$ is given by 
\begin{equation}
\label{eq-delVdelx}
    \frac{\dd V_{\mr{CW}}}{\dd x}
    =
    \frac{m_0^2 M_Q^2}{16\pi^2} x^{k-1} e^{-px}
        (k-px)
        \left(
            \log{\frac{M_Q^2}{\mu^2}+k\log{x}-px}
        \right) + \mc{O}(|q|).
\end{equation}
This has zero points at $x = k/p$ and 
\begin{align}
\label{eq-MinLoc}
    x = x_0 := -\frac{k}{p} \Wcal_{-1}\left(
     -\frac{p}{k} \left(\frac{\mu}{M_Q}\right)^{2/k}
     \right),  
\end{align}
where $\Wcal_{-1}(z)$ is the Lambert function 
satisfying $\Wcal_{-1}(z) e^{\Wcal_{-1}(z)} = z$  
and is on the branch with $\Wcal(z) < -1$ at $-e^{-1} < z < 0$. 
The former $x=k/p$ is a maximum, and the latter $x=x_0$ is the minimum where the modulus is stabilized. 
At this minimum, the renormalization scale $\mu$ 
determines the vector-like quark mass at the minimum, 
\begin{align}
\label{eq:potential_minimum}
\mu^2 = m_Q^2(x_0) \simeq M_Q^2 x_0^k e^{-p x_0}. 
\end{align}

The CW potential would realize the inflation at large $x$. 
As the potential has negative value, 
we uplift the potential by a constant term, 
so that the cosmological constant is almost vanishing. 
The full potential is given by 
\begin{align}
    V(x) := V_{\CW}(x) - V_{\CW}(x_0). 
\end{align}
The constant term is approximately given by 
\begin{align}
    V_{\CW}(x_0) = - \frac{m_0^2 m_Q^2(x_0)}{16\pi^2} 
                  \left[ 1 - 
                   \frac{m_0^4}{6 m_Q^4(x_0)} 
                   + \order{\frac{m_0^6}{m_Q^6(x_0)}} \right]. 
\end{align}
Figure~\ref{f_CWpotential} 
shows the normalized potential shape,  
where the modular form is $Y_{1'}^{(12)}$ of $\Gamma_3$, 
$m_0/M_Q = 10^{-5}$ with various $\mu/M_Q$.
We find that there is an exponentially flat plateau 
at $x \gg 1$, which could realize the inflation. 
We shall focus on $t>0$ because 
the exponential plateau is realized only in this case. 
We note that the constant uplifting 
is crucial to have the plateau. 
Such constant uplifting would be realized 
by the F-term uplifting~\cite{Lebedev:2006qq,Abe:2006xp,Dudas:2006gr,Gomez-Reino:2006tjy,Kallosh:2006dv,Abe:2007yb}~\footnote{
For instance, if there is an additional singlet field $X$ 
with weight $-h$ and having the following terms, 
\begin{align}
    \Delta K = \frac{\abs{X}^2}{(-i\tau+i\tau^\dag)^h}, 
    \quad 
    \Delta W = \xi_X X, 
\end{align}
the F-term potential becomes the constant uplifting term, 
$\Delta V = \abs{\xi_X}^2$, where $\xi_X$ is independent of $\tau$, although it may depend on other moduli when the model is embedded into string theory.
Further discussions, such as the stabilization of $X$, 
are beyond the scope of this paper. }.

\subsection{Slow-roll parameters and observables}
\label{sec-slowroll}

We discuss the inflation driven by the CW potential. 
We define the canonically normalized fields of the modulus $\tau$ as 
\begin{align}
\label{eq-cnbase}
     \phi := \sqrt{\frac{h}{2}} \log\left(\frac{x}{2}\right), 
     \quad 
     a := \sqrt{2h}\; \frac{\rta}{x}, 
\end{align}
so that the kinetic terms are induced from the \kahler potential in Eq.~\eqref{eq:model_Lagrangian}
\begin{align}
\label{eq-Lkin}
   \mathcal{L}_{\mathrm{kin}} 
   = \frac{h}{(2\ita)^2} \left[ (\partial_\mu \ita)^2 + 
                                 (\partial_\mu \rta)^2 \right] 
   = \frac{1}{2} \left(\partial_\mu \phi\right)^2 
     + \frac{1}{2} \left(\partial_\mu a \right)^2.  
\end{align}
We identify $\phi$ as the inflaton and $a$ as the axion\footnote{
Here, the axion is not the QCD axion since the CW potential 
dominates over the QCD contribution. 
Nevertheless, we name the physical field along $\rta$ as the axion, as the CP-odd partner of the moduli.
}. 
Since the axion is irrelevant for the inflationary dynamics, 
we only consider its canonically normalized form around the vacuum.
The physics of the axion will be discussed in the following sections.

To proceed, we define the rational quantities as  
\begin{align}
    F(x) := \frac{V(x)}{V_0}, 
    \quad 
    y(x) := \frac{m_Q^2(x)}{\mu^2}, 
    \quad 
    z := \frac{m_0^2}{\mu^2},
\end{align}
where $V_0 := m_0^2\mu^2/(16\pi^2)$, and note that $\mu$ satisfy Eq.~\eqref{eq:potential_minimum}. 
Here, $x$ is understood as the function of $\phi$. 
Note that $y\to 1$ for $x\to x_0$ 
and $y \to 0$ for $x\to \infty$. 
Hence, $z/y \ll 1$ near the minimum, 
while it can be sizable for large $x$. 
Nonetheless, we assume $z/y \ll 1$ for our analytical estimation.  
The normalized potential $F$ is arranged to
\begin{align}
F(x)  =&\ 
  \frac{y^2}{2z} \left[
    \left(1+\frac{z}{y}\right)^2\left( 
    \log (y+z) - \frac{3}{2}
    \right) - \left(\log y -\frac{3}{2}\right) 
    \right]
    \\ \notag 
    &\ \hspace{2.5cm}
    -\frac{1}{2z} \left[
    \left(1+z\right)^2\left( 
    \log (1+z) - \frac{3}{2}
    \right) +\frac{3}{2}
    \right]
    \nn \\
\label{eq-Fexact}
    =&\ 1 + y(\log y - 1) +\frac{z}{2} \log y 
          - z \sum_{n=1}^\infty \frac{(-z)^n}{n(n+1)(n+2)} 
               \left(\frac{1}{y^n}-1\right). 
\end{align}
The derivatives with respect to $y$ are given by 
\begin{align}
    F_y = \log y - 1 + \left( 1 + \frac{y}{z} \right) \log \left( 1 + \frac{z}{y} \right),
    \quad 
    F_{yy} = \frac{1}{z} \log \left( 1 + \frac{z}{y} \right),
    \quad 
    F_{yyy} = - \frac{1}{y(y+z)}.
\end{align}
Here, we denote $F_y := \partial_y F$, and similarly for the other derivatives. 
It is straightforward to obtain the exact formula 
of the derivatives with respect to $\phi$, which are shown in App.~\ref{app:derivatives}. 
The asymptotic behavior of the normalized potential 
at $z \ll y \ll 1$ is given by 
\begin{align}
\label{eq-Fasym}
    F \sim 1- 2pe^{px_0} \left(\frac{2}{x_0}\right)^k \mathrm{exp}\left[ 
    (k+1)\sqrt{\frac{2}{h}} \phi - 2pe^{\sqrt{\frac{2}{h}} \phi}
     \right] \left( 1 + \order{\frac{k}{p}\phi e^{-\sqrt{\frac{2}{h}} \phi} }\right). 
\end{align}
Thus, the flatness of the inflationary plateau is ensured by the double-exponential behavior of the potential in terms of $\phi$ \cite{Abe:2023ylh}.
In the regime $\abs{q}\ll 1$,
this behavior originates from the asymptotic form
$\Ykr{k}{1_t}\sim q^{t/N}\sim \exp(-p e^{\phi})$, where we set $h=2$ to simplify the notation.
Hence, a double-exponentially flat potential is realized at $\ita\gg 1$,
provided that the leading potential is generated by a non-trivial singlet modular form.
Note that the condition $\ita\gg 1$ is also the regime in which the K\"ahler potential with the rescaling symmetry
in Eq.~\eqref{eq:model_Lagrangian} is expected to arise in string compactifications.
As discussed in the next section, the same regime also gives rise to a light axionic state
associated with the approximate shift symmetry $\tau\to\tau+c$, where $c\in\mathbb{R}$.

Now we calculate the inflationary observables on the CW potential. 
The slow-roll parameters are given by
\begin{equation}
    \epsilon_V=\frac{1}{2}\left( \frac{F_\phi}{F} \right)^2,
    \quad
    \eta_V = \frac{F_{\phi\phi}}{F},
    \label{eq:slow-roll-parameters}
\end{equation}
where the index ${\phi}$ represents the derivative with respect to $\phi$. 
To see the qualitative behavior, we assume $z \ll y \ll 1$, so that the slow-roll parameters are approximately given by 
\begin{align}
\label{eq-appEpsEta}
   \epsilon_V \simeq
   \frac{1}{h} \left(px\right)^2 
   \left(y\log y\right)^2, 
\quad 
   \eta_V \simeq
   \frac{2}{h} \left(px\right)^2 y \log y, 
\end{align}
where $k/px \ll 1 \ll |\log y |$ is also assumed. 
Hence, the slow-roll parameters are suppressed by $y \ll 1$ as 
$\eps_V \sim \order{(y\log y)^2}$ and $\eta_V \sim \order{y\log y}$. 
Since the powers in $y$ are different, 
$\eps_V \ll |\eta_V| $ is expected from the approximate forms.

The e-folding number is evaluated as 
\begin{align}
    N_e = \int^{\phi_*}_{\phi_f} \frac{F}{F_\phi} d\phi 
        = \frac{h}{2} \int^{x_*}_{x_f} 
           \frac{1+y(\log y -1)}{ y \log y } \frac{dx}{(k-px)x},   
\end{align}
where $\phi_{f}$ ($x_f$) and $\phi_{*}$ ($x_*$) 
are the values of $\phi$ ($x$) at the end of inflation 
and at the horizon exit, respectively. 
We take the end of inflation at $\max(\eps_V, \abs{\eta_V}) = 1$, 
where the slow-roll condition is violated. 
The time of horizon exit is $N_e \sim 50\mathrm{-}60$ before the end of inflation. 
By further assuming $px_f \gg k$ and $x_* \gg x_0$,
the e-folding number is estimated as 
\begin{align}
    N_e \sim 
        \frac{h}{2p^2} x_0^k e^{-px_0}
        \int^{x_*}_{x_f} dx \frac{e^{px}}{x^{k+3}}
         \sim     \frac{h}{2(px_*)^3y_*} . 
         \left[1 + \order{\frac{x_0}{x_f}} \right]. 
\end{align}
Thus, $N_e \sim 50\mathrm{-}60$ is realized when 
\begin{align}
y_*
= \frac{m_Q^2(x_*)}{m_Q^2(x_0)} 
\sim  \frac{h}{2(px_*)^3 N_e}, 
\end{align}
so that, for $p x_*=\mathcal O(10)$, 
the vector-like quark mass squared at the pivot scale is roughly four orders of magnitude smaller than that at the minimum.

The power spectrum is estimated as 
\begin{align}
 \label{eq:PWR_spectrum}
 \mathcal{P}_R =&\ \frac{V}{24\pi^2 \eps_V}
               \sim \frac{hV_0}{24\pi^2 (px_*)^2 (y_*\log y_*)^2} 
                     \left(1+\order{y_*, \frac{k}{px_*}}\right), 
\end{align}
where the right-hand side is evaluated at $\phi =\phi_*$. 
We fix the soft mass squared $m^2_0$ to realize the Planck data~\cite{Planck:2018vyg,Planck:2018jri}. 
The ratio $z := m_0^2/\mu^2$ is estimated as 
\begin{align}
\label{eq-detz}
    z \sim&\ \frac{384\pi^4}{h\mu^4}  
           \left(px_*\right)^2
            \left(y_*\log y_*\right)^2 \Pcal_R 
            \\ \notag 
      \sim&\ 7.9\times 10^{-7} \times 
           \left(\frac{0.1}{\mu}\right)^4 
           \left(\frac{p x_* }{10}\right)^2 
           \left(\frac{y_*\log y_*}{10^{-4}}\right)^2
           \left(\frac{\Pcal_R}{2.1\times 10^{-9}}\right),  
\end{align}
where $\mu$ is in the reduced Planck unit $M_p = 1$. 
Note that the approximations, such as $z \ll y \ll 1$ and $x_* \gg x_0$,  
are not quantitatively precise when $N_e$ and $\Pcal_R$ are fit to the observed data, and only describe the qualitative behaviors. 
We shall rely on the numerical evaluation in the following analysis.

\begin{table}[t]
   \centering
   \begin{tabular}{lllc} \hline
      Observable & Value & {} &Refs. \\ \hline\hline
      $\mc{P}_{\mc{R}}\times10^9$ & $2.10\pm0.030$
      & (TT,TE,EE+lowE+lensing) & \cite{Planck:2018vyg,Planck:2018jri}
      \\
      \hline
      $n_s$ & $0.965\pm0.0042$
      &$\text{(TT,TE,EE+lowE+lensing)}$ & \cite{Planck:2018vyg,Planck:2018jri}
      \\ 
      {} & $0.974\pm0.0034$ & $\text{(Planck+ACT+LB)}$ & \cite{ACT:2025fju}
      \\
      \hline
      $n_s$ & $0.964\pm0.0044$
      & {} & {}
      \\
      $\alpha_s$ & $-0.0045\pm0.0067$
      & $\text{(TT,TE,EE+lowE+lensing)}$ & \cite{Planck:2018vyg,Planck:2018jri}
      \\
      \hline
      $r$ & $<0.032$ & (Planck+BK18+BAO) & \cite{Tristram:2021tvh}
      \\
      \hline
   \end{tabular}
   \caption{
        Cosmological parameters used for comparison with inflationary predictions.
        The entries in parentheses indicate either
        the combination of observational channels used for Planck-only results,
        or the experiments and datasets used in the combined analyses.
    }
   \label{t:ObservationalResults}
\end{table}

Finally, the spectral index $n_s$,
the running spectral index $\alpha_s$, and the tensor-to-scalar ratio $r$ 
are expressed by the slow-roll parameters as~\cite{Baumann:2009ds} 
\begin{align}
    n_{s}  
    =1-6\epsilon_V + 2\eta_V, 
    \quad 
    \alpha_s = 16\epsilon_V \, \eta_V - 24 \epsilon_V^2 - 2\xi^2,
    \quad 
    r 
    = 16\epsilon_V, 
    \label{eq:tts-ratio}
\end{align}
where $\xi$ is defined as 
\begin{equation}
    \xi^2 \coloneqq \frac{F_\phi F_{\phi\phi\phi}}{F^2} 
    \sim \frac{4}{h^2} \left(p x\right)^4 \left(y\log y\right)^2. 
\end{equation}
Here, we have used the same approximation as Eq.~\eqref{eq-appEpsEta} to obtain the last equality.
Their observed values are summarized in Table~\ref{t:ObservationalResults}.
Since $\eps_V < \xi^2 \ll |\eta_V| \ll 1$, 
these are approximately given by, 
\begin{align}
    n_s -1 \sim  -\frac{4}{h} (px)^2 y \abs{\log y}, 
    \quad 
    \alpha_s \sim -\frac{8}{h^2} (px)^4 (y\log y)^2, 
    \quad  
    r \sim \frac{16}{h} (px)^2 (y\log y)^2. 
\end{align}
Thus, the tensor-to-scalar ratio $r$ is expected to be very small. 
The running spectral index $\alpha_s$ is also small, 
but two orders of magnitude larger than $r$ due to the factor $(px)^4$.

\subsection{Numerical analysis}
\label{sec:NumericalAnalysis}

\begin{figure}[htbp]
\centering

\begin{subfigure}{0.48\textwidth}
\centering
\includegraphics[width=\linewidth]{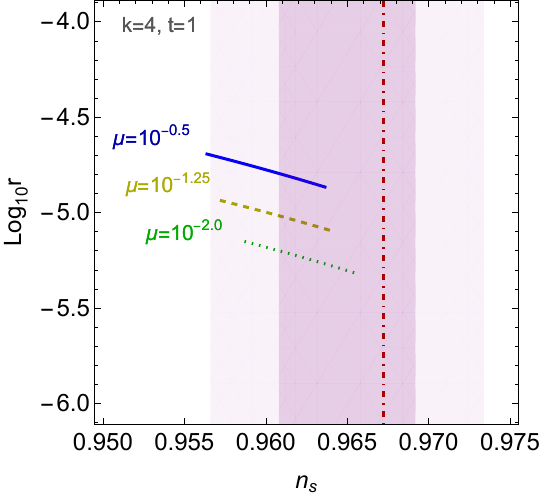}
\end{subfigure}
\hfill
\begin{subfigure}{0.48\textwidth}
\centering
\includegraphics[width=\linewidth]{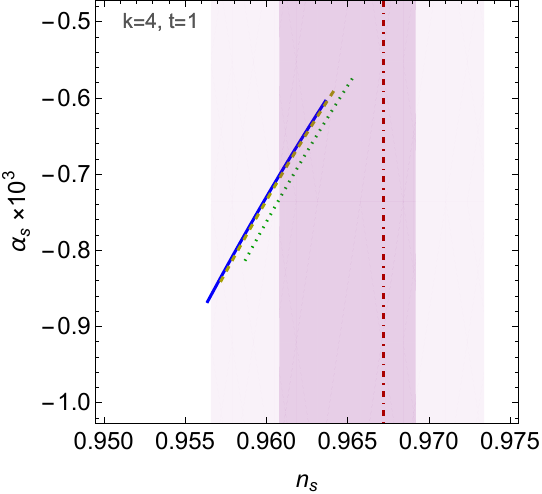}
\end{subfigure}

\vspace{0.5em}

\begin{subfigure}{0.48\textwidth}
\centering
\includegraphics[width=\linewidth]{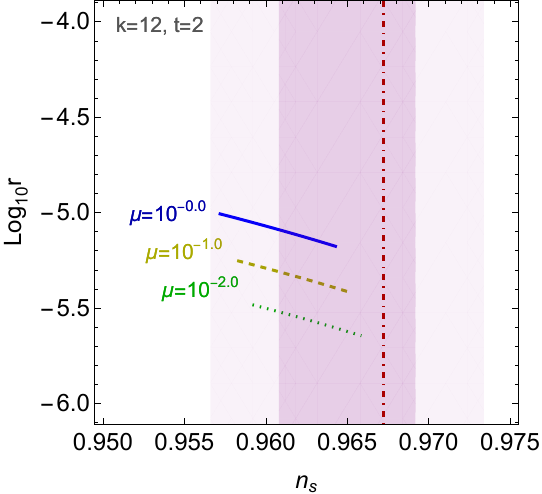}
\end{subfigure}
\hfill
\begin{subfigure}{0.48\textwidth}
\centering
\includegraphics[width=\linewidth]{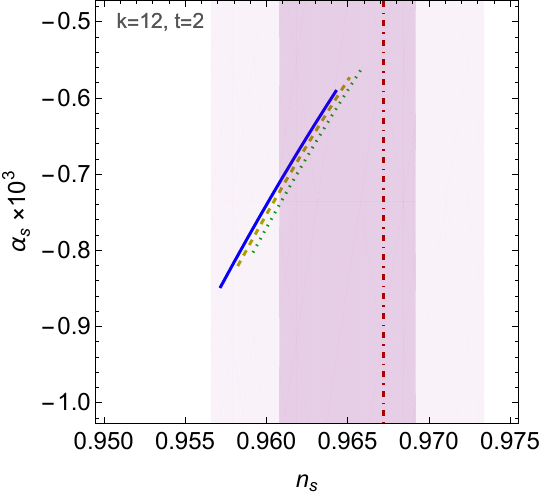}
\end{subfigure}
\caption{\label{fig-nsra}
Inflationary observables in the CW potential.
The left panels show the predictions in the $(n_s, r)$ plane,
while the right panels show those in the $(n_s, \alpha_s)$ plane.
The upper panels correspond to $(k,t)=(4,1)$ and the lower panels to $(k,t)=(12,2)$ for the modular form $Y^{(k)}_{1_t}$.
Each line corresponds to a different value of the renormalization scale $\mu$ in Planck units.
The left ends of the lines correspond to $N_e = 50$,
and $N_e$ increases to the right up to 60.
The purple region indicates the allowed range from Planck observations~\cite{Planck:2018jri,Planck:2018vyg},
and the red dot-dashed line is the $2\sigma$ lower bound 
from the ACT data~\cite{ACT:2025fju}.
}
\end{figure}

\begin{table}[t]
    \centering
     \begin{tabular}{c|cccc} \hline
  & BP1 & BP2 & BP3 & BP4 \\ \hline\hline
$(t,k)$ & (1,4) & (1,4) & (2,12) & (2,12) \\ 
$\log_{10}\mu/M_p$ & -0.5 & -2.5 & 0.0 & -2.0 \\ 
$m_0\times 10^{-14}$ [GeV] & 0.681 & 12.848 & 0.15 & 8.759 \\ \hline
$x_0$ & 3.482 & 8.48 & 3.895 & 8.24 \\ 
$x_*$ & 11.378 & 15.42 & 9.42 & 12.544 \\ 
$x_f$ & 8.075 & 12.928 & 7.514 & 11.159 \\ \hline
$n_s$ & 0.96 & 0.962 & 0.961 & 0.962 \\ 
$r\times 10^6$ & 16.422 & 5.776 & 8.006 & 2.698 \\ 
$-\alpha_s\times 10^4$ & 7.185 & 6.753 & 7.027 & 6.663 \\ \hline
$m_\phi \times 10^{-12}$ [GeV] & 7.987 & 19.898 & 7.313 & 22.199 \\ 
$T_R \times 10^{-9}$ [GeV] & 1.111 & 4.369 & 0.973 & 5.148 \\ \hline
$f_a\times 10^{-17}$ [GeV] & 1.551 & 0.636 & 1.386 & 0.655 \\ 
$m_a$ [PeV] & 361.296 & 0.266 & 1188.31 & 1.995 \\ 
$T_d$ [MeV] & 26658.5 & 1.3 & 177898. & 25.899 \\ \hline
$\mathcal{P}_{\zeta_a}/\mathcal{P}_{\mathcal{R}}$ & 0.021 & 0.045 & \
0.013 & 0.02 \\ 
$f_{\mathrm{NL}}^a\times 10^{4}$ & -5.953 & -25.943 & -2.152 & -5.05 \
\\ 
$\Delta \times 10^{-3}$ & 0.58 & 7890.67 & 0.06 & 494.279 \\ \hline
\hline 
    \end{tabular}
    \caption{
     Values of the parameters and inflationary observables 
     at the benchmark points (BPs) 
     under the finite modular symmetry $\Gamma_3$.
     The e-folding number is set to $N_e = 55$ at all the BPs. 
     To calculate the axion observables, see Sec.~\ref{sec:axion-cos} for quantities related to the axion, 
     we set $m_f = m_{\tilde f} = 10^{14}\,\GeV$, $\theta_i = 0.1$, $h=1$, $c_G = 1$, $\sum_G N_G = 12$, 
     and $\alpha_G = 1/24$, $k_G=12$ for all the BPs. }
     \label{tab:BPs}
\end{table}

We numerically demonstrate the inflation driven by the CW potential,
whose shape is given by Eq.~\eqref{eq-Fexact}. 
For illustration, we consider $\Gamma_3 \simeq A_4$ 
and two modular forms $Y^{(4)}_{1_1}$ and $Y^{(12)}_{1_2}$. 
We note that the 
choice of the weight $k$ and representation $r$ affect to the inflationary observables 
only quantitatively, 
whereas the renormalization scale $\mu$ change the values significantly through Eq.~\eqref{eq-detz}. 
Indeed, we will see the two cases show similar results, 
and hence the results will be similar for the other choices of the modular form. 
As we are interested in the region where $x := 2\ita$ is large, 
we only keep the leading term in the $q$-expansion, 
i.e., $Y^{(k)}_{1_t} = q^{t/3} (1+ \order{\abs{q}})$ 
with $\abs{q} = e^{-\pi x} \ll 1$. 
We take the scale of the bare VLQ mass parameter 
in the superpotential at $M_Q = M_p$. 
The value of the soft SUSY breaking mass $m_0$ 
is fixed by Eq.~\eqref{eq:PWR_spectrum}, 
such that the power spectrum $\Pcal_R$ is fit to the observed value.

With the above conditions, 
there remains only one parameter $\mu$,
corresponding to the VLQ mass at the minimum. 
The value of $\mu$ should be sufficiently small 
to have the minimum in Eq.~\eqref{eq-MinLoc}, 
i.e. the argument of the Lambert function is larger than $-e^{-1}$. 
On the other hand, it should be large enough to explain the power spectrum 
while keeping $z/y = m_0^2/m^2_Q(x) \ll 1$. 
Therefore, we find that the scale should be of order
$\mu \sim 10^{-2}\text{--}10^0$
in the Planck unit.

Figure~\ref{fig-nsra} shows the inflationary observables in our model. 
The left panels show the plots on the $(n_s, r)$ plane, 
and the right ones show those on the $(n_s, \alpha_s)$ plane. 
The upper panels correspond to the modular form $Y^{(k)}_{1_t}$ with $(k,t)=(4,1)$, while the lower panels correspond to $(k,t)=(12,2)$.
Each line corresponds to a different value of the scale $\mu$ in the Planck unit. 
The left end of each curve corresponds to \(N_e = 50\), and \(N_e\) increases along the curve toward the right, reaching \(N_e = 60\) at the right end in all panels.
The darker (lighter) purple region 
is the $1\sigma$ ($2\sigma$) allowed range according to the Planck data. 
The upper bound on the tensor-to-scalar ratio $r < 0.032$ 
is a few orders of magnitude larger than the predictions. 
The running spectral index $\alpha_s$ is consistent with zero, 
and the prediction is an order of magnitude smaller than the bound. 
Thus $r$ and $\alpha_s$ are much smaller than the current sensitivities.

The spectral index is significantly tilted away from scale invariance, yielding \(n_s \simeq 0.965\), because \(\eta_V \sim y\log y\) dominates over the other slow-roll parameters \(\eps_V\) and \(\xi\).
In Fig.~\ref{fig-nsra}, 
the purple region indicates the allowed range from Planck observations~\cite{Planck:2018jri,Planck:2018vyg},
and the red dot-dashed line is the $2\sigma$ lower bound 
from the ACT data~\cite{ACT:2025fju}.
Hence, the CW prediction predicts $n_s$ consistent 
with the Planck data, whereas the value is slightly smaller than the ACT result 
$n_s = 0.974\pm 0.0034$.
Thus, if the current hints from ACT are confirmed, future observations will be able 
to test the model~\footnote{See also Ref.~\cite{Kallosh:2025rni}.}.

Table~\ref{tab:BPs} summarizes the values 
of the input parameters and the observables. 
The value of the soft SUSY breaking mass $m_0$ 
is fixed to realize the power spectrum $\Pcal_R$. 
$x_0$, $x_f$, and $x_*$ correspond to the position of the minimum, 
the end of inflation determined by the violation of the slow-roll condition $\abs{\eta_V}<1$, 
and the horizon exit value at $N_e=55$, respectively. 
As expected from the analytical calculation in Eq.~\eqref{eq-appEpsEta}, 
$\eps_V \ll \abs{\eta_V}$ holds, and hence the slow-roll condition 
is violated by $\eta_V$ rather than $\eps_V$. 
The spectral index $n_s$ is consistent with the Planck data, 
and both $r$ and $\alpha_s$ are much smaller than 
the current sensitivities. 
Therefore, 
we conclude that the CW potential of the modulus $\phi$ 
realizes realistic inflation. 
The observables related to the inflaton decay and the axion 
will be discussed in the following sections.

It is intriguing that the favored values $\ita_0 = x_0/2$ are 
in the suitable range to explain the mass hierarchies of the quarks and leptons 
by the Froggatt-Nielsen (FN) mechanism of the residual $\mathbb{Z}^T_3$ symmetry. 
The hierarchy parameter is $q_{\FN}^{1/3} \sim 0.015$ for $x_0 \sim 4$, 
which are realized at BP1 and BP3. 
Whereas, $q^{1/3}_{\FN} \sim 0.0002$ at BP2 and BP4. 
These values would be used to partially explain 
the quark and lepton hierarchies in the SM. 
Note that the automorphic factor $(2\ita)^{k_\Phi}$ 
is also important to determine the hierarchies~\cite{Petcov:2023vws,Abe:2023qmr,Petcov:2026mdx}. 
Thus, we see the potential correlation between the inflaton and the flavor structure, 
that the smaller flavor hierarchy by $q_{\FN}^{1/N}$ 
favors a smaller spectral index $n_s$ as read from Table~\ref{tab:BPs}. 
The further discussion is beyond the scope of this paper.

\section{Reheating after inflation}
\label{sec:reheating}

We now discuss the reheating after inflation.
In order for the inflaton to decay predominantly into gauge bosons, we assume that the gauge kinetic function $f_G$ depends on the modulus $\tau$
\begin{align}
\label{eq-gaugekinf}
    f_G(\tau) = f_0 + \frac{1}{16\pi^2} f_{1G}(\tau), 
\end{align}
where $f_1(\tau)$ is a holomorphic function of $\tau$. 
The index $G$ runs over the SM gauge group, 
and we assume that the tree-level value $f_0$ is universal to $G$ 
motivated by the gauge coupling unification of the MSSM. 
In Refs.~\cite{Kaplunovsky:1987rp,Dixon:1990pc,Feruglio:2023uof,Feruglio:2024ytl,Feruglio:2025ajb}, 
this is identified as 
\begin{align}
    f_{1G}(\tau) = \log \eta(\tau)^{2k_G}.
\end{align}
Here, $\eta(\tau)$ is the Dedekind eta function,  
so that the anomaly of the modular transformation 
is compensated by the shift of $f_1(\tau)$ \cite{Derendinger:1991hq,Ibanez:1992hc}. 
The qualitative results of reheating are expected to be insensitive to this choice.
A value of the integer $k_G$ depends on the transformation properties  
of the MSSM fields for the anomaly cancellation. 
In the following, we assume this form of the gauge kinetic term 
and treat the integer $k_G$ as a parameter.

At the minimum of the CW potential, 
the inflaton mass is estimated as 
\begin{align}
    m_\phi =&\ \frac{m_0}{\sqrt{8\pi^2 h}} \abs{k-px_0} \frac{\mu}{M_p} 
    \\ \notag 
    \sim&\ 4.7\times 10^{11}~\GeV
           \times\frac{1}{\sqrt{h}} 
           \left(\frac{m_0}{10^{14}~\GeV}\right) 
           \left(\frac{\abs{k-px_0}}{10}\right)
           \left(\frac{\mu/M_p}{10^{-2.5}}\right). 
\end{align}
Here and hereafter, we neglect the $\order{z}$ corrections 
since they are always tiny near the minimum, see Table~\ref{tab:BPs}. 
Due to the suppression by the loop factor and $M_p$, 
the inflaton mass $m_\phi$ is smaller 
than the soft SUSY breaking mass $m_0$ and the VLQ mass $m_Q(x_0) = \mu$. 
Thus, the MSSM superpartners and the vector-like quarks are heavier than the inflaton, 
and the inflaton does not decay into them.

The decay widths of the inflaton $\phi$ 
to the gauge boson $V$ and the axion $a$, 
defined in Eq.~\eqref{eq-cnbase}, 
are given by
\begin{align}
    \Gamma_{\phi\to VV} =&\ \sum_G \frac{N_G}{294912\pi^3 h} 
                      \left(\frac{k_G x_0}{\mathrm{Re}f_G}\right)^2 
                      \frac{m_\phi^3}{M_p^2} 
    \\ \notag 
     \sim&\ 0.52~\MeV \times \left(\frac{N_G}{12h}\right)  
                      \left(\frac{2.0}{\mathrm{Re}f_G}\right)^2 
                      \left(\frac{k_G}{12}\right)^2 
                      \left(\frac{x_0}{8.0}\right)^2 
                      \left(\frac{m_\phi}{10^{12}~\GeV}\right)^3,  
                      \\ 
   \Gamma_{\phi\to aa} =&\ \frac{1}{16\pi h} \frac{m_\phi^3}{M_p^2} 
   \sim 
   3.5~\MeV \times  \frac{1}{h}
   \left(\frac{m_\phi}{10^{12}~\GeV}\right)^3
   \label{eq-Gammaa},
\end{align}
where $N_G = 1,3$ and $8$ is the number of gauge bosons 
of $U(1)_Y$, $SU(2)_L$ and $SU(3)_C$, respectively. 
Here, \(\mathrm{Re}f_G \simeq \mathrm{Re} f_0 = 2\) corresponds to the MSSM gauge coupling, with \(\alpha_G^{-1} \sim 25\) at the coupling unification scale.
These two decay modes dominate over all other decay channels.

The total width of the inflaton is approximately given by 
\begin{align}
\label{eq-GamPhi}
    \Gamma_\phi \simeq \Gamma_{\phi\to VV} + \Gamma_{\phi\to aa}
    =: \frac{c_G }{16\pi h } \frac{m_\phi^3}{M_p^2},
    \quad 
    c_G := 1 
      + \frac{3x_0^2}{128\pi^2} 
      \left(\frac{2.0}{\mathrm{Re}f_G}\right)^2 
      \left(\frac{k_G}{12}\right)^2,
\end{align}
where $\sum_{G} N_G = 12$ and $k_G$ are assumed to be universal. 
Thus, the reheating temperature $T_R$ via the inflaton decay reads 
\begin{align}
\label{eq-TR}
    T_R =&\ (1-B_a)^{\frac{1}{4}} 
           \left(\frac{90}{\pi^2 g_*(T_R)}\right)^{\frac{1}{4}}
           \sqrt{\Gamma_\phi M_p} 
           \\ \notag 
       \sim&\ 1.5\times 10^{8}~\GeV 
       \times (1-B_a)^{\frac{1}{4}} \sqrt{\frac{c_G}{h}} 
          \left(\frac{106.75}{g_*(T_R)}\right)^{\frac{1}{4}} 
          \left(\frac{m_\phi}{10^{12}~\GeV}\right)^{\frac{3}{2}}, 
          \label{eq:TR}
\end{align}
where $B_a := \Gamma_{\phi\to aa}/\Gamma_\phi$ is the branching fraction to the axions.
Here, $g_*(T_R)$ denotes the effective number of relativistic degrees of freedom at reheating. 
Note that the axions produced from inflaton decay are independent of the SM thermal bath temperature due to their feeble interactions, and hence give rise to the factor $(1-B_a)^{1/4}$.
We see that the reheating occurs well before the Big-Bang Nucleosynthesis (BBN). 
As discussed in the next section, axions decay into SM particles before BBN, 
and hence those produced via inflaton decay do not lead to the dark radiation.
Alternatively, the branching fraction into axions can be suppressed by additional interactions allowed in the K\"ahler potential, as proposed in Refs.~\cite{Cicoli:2012aq,Higaki:2012ar,Higaki:2013lra}.

\section{Axion cosmology}
\label{sec:axion-cos}

We discuss the role of the axion in the model. We show that the energy density from axion oscillations dominates over the radiation energy density and that the axion decays before BBN. 
As discussed in App.~\ref{app:axion-abundance} in detail,  
the axions produced via inflaton decay are relativistic at reheating and subsequently redshift as radiation, 
eventually becoming non-relativistic as the Universe expands. 
Their abundance depends on the inflaton branching fraction 
but is typically subdominant compared to that from axion oscillations 
via the misalignment mechanism in the relevant parameter space. 
We therefore neglect their contribution in the following analysis.
We also argue that curvature perturbations sourced by axion oscillations would induce isocurvature perturbations.

\subsection{Axion mass and decays}
\label{sec:axion}

The axion, the CP partner of the inflaton $\phi$, plays an important role after inflaton decay. As discussed in Ref.~\cite{Higaki:2024ueb}, the axion is massless at the minimum of the CW potential, where there is only one VLQ pair. 
The case with a second VLQ is discussed in Ref.~\cite{Higaki:2024jdk}, 
while the case with the SM fermion masses is discussed in Ref.~\cite{Higaki:2024ueb}.
The CW potential along $\rta$ is schematically given by 
\begin{align}
    V_\CW(\rta) \sim c_1 \frac{m_f^2m_{\tilde{f}}^2}{16\pi^2} \abs{q} 
                  \cos\left(\frac{a}{f_a}\right) 
                  =: \Lambda_{\CW}^4 \cos\left(\frac{a}{f_a}\right), 
\end{align}
where $m_f$ ($m_{\tilde{f}}$) is a fermion (sfermion) 
mass whose mass depends on the modulus $\tau$. 
Here $c_1$ is the coefficient of the $q$-expansion 
of the modular form, defined in Eq.~\eqref{eq:qexpansion}. 
The axion decay constant $f_a$ is given by  
\begin{align}
    f_a = \sqrt{\frac{h}{2}} \frac{M_p}{2\pi \ita_0} 
        \sim 5.0\times 10^{16}~\GeV \times \sqrt{h} \left(\frac{2.0}{\ita_0}\right), 
\end{align}
after the canonical normalization in Eq.~\eqref{eq-cnbase}. 
The scale of the CW potential for the axion is estimated as 
\begin{align}
    \Lambda_{\CW} \sim 8.9\times 10^{9}~\GeV \times  
             c_1^{\frac{1}{4}}
             \left(\frac{m_f}{10^{10}~\GeV}\right)^{\frac{1}{2}}
             \left(\frac{m_{\tilde{f}}}{10^{14}~\GeV}\right)^{\frac{1}{2}} 
             \left(\frac{\abs{q}^{1/3}}{10^{-2}}\right)^{3/4}. 
\end{align}
This scale is predominantly given 
by a supermultiplet whose $m_f m_{\tilde{f}}$ is the largest. 
Here, we have assumed $m_{\tilde{f}} \sim m_0$ and 
$m_f \ll \mu$ 
so that the potential associated with axion does not affect the inflationary trajectory in the model.
Thus, the typical size of the CW potential for the axion 
tends to be much higher than the QCD potential, 
so the axion originating from the modulus $\tau$ is not the QCD axion. 
The axion mass is estimated as 
\begin{align}
    m_a \simeq &\ \frac{m_f m_{\tilde{f}}}{2\sqrt{2}\pi f_a} 
            \sqrt{\abs{c_1 L_2}} e^{-\pi x_0/2}
        \\ \notag     
        \sim&\ 140~\TeV \times 
             \left(\frac{\abs{c_1 L_2}}{100}\right)^{\frac{1}{2}}
             \left(\frac{m_f}{10^{10}~\GeV}\right)
             \left(\frac{m_{\tilde{f}}}{10^{14}~\GeV}\right) 
             \left(\frac{10^{16}~\GeV}{f_a}\right) 
             \left(\frac{\abs{q}^{1/3}}{10^{-3}}\right)^\frac{3}{2},
\end{align}
where 
\begin{align}
  L_2 := \log \frac{m_f^2}{\mu^2} - 1 
        + \left(1+\frac{m_f^2}{m_{\tilde{f}}^2}\right) 
         \log \left(1+\frac{m_{\tilde{f}}^2}{m_f^2}\right) 
      \sim 
      \begin{cases}
          \log \dfrac{m_f^2}{\mu^2} & m_{\tilde{f}}^2 \ll m_f^2 \\ 
          \log \dfrac{m_{\tilde{f}}^2}{\mu^2} -1  
                & m_f^2 \ll m_{\tilde{f}}^2 \
      \end{cases}. 
\end{align}
Therefore, the axion $a$ is much lighter than the modulus $\phi$. 
This can be understood by the approximate continuous shift symmetry $\tau \to \tau + c$ 
which is explicitly broken by $\order{\abs{q}}$ corrections~\cite{Higaki:2024jdk}.

The axion may decay through the coupling induced by the gauge kinetic function in Eq.~\eqref{eq-gaugekinf}. 
It predominantly decays into SM gauge bosons, and its decay width is given by
\begin{align}
    \Gamma_a \simeq \Gamma_{a\to VV} 
     = \sum_G \frac{N_G \alpha_G^2}{4096\pi^3} \left(\frac{k_G}{12}\right)^2 \frac{m_a^3}{f_a^2}.
\end{align}
The decay temperature reads 
\begin{align}
\label{eq-Td}
    T_d =&\ \left(\frac{90}{\pi^2 g_*(T_d)}\right)^\frac{1}{4} 
           \sqrt{M_p \Gamma_{a}} 
           \\ \notag 
        \sim&\ 1.3~\MeV\times \left(\frac{g_*(T_d)}{10.75}\right)^\frac{1}{4}
             \left(\frac{\sum_G\sqrt{N_G}\alpha_G k_G}{1.2}\right) 
             \left(\frac{10^{16}~\GeV}{f_a}\right) 
             \left(\frac{m_a}{100~\TeV}\right)^\frac{3}{2}. 
\end{align}
Thus, the axion decays before the BBN if it is heavier than $\order{100~\TeV}$.

\subsection{Cosmological role of the axion} 

The axion remains light compared to the Hubble scale $H$ during inflation 
and hence acquires sizable quantum fluctuations of its own.
After inflation, as the Hubble parameter decreases and becomes comparable to the axion mass, the axion begins coherent oscillations around its minimum. The oscillation amplitude is determined by the initial misalignment of the axion $a_i=f_a\theta_i$. The resulting oscillation energy may subsequently dominate the Universe after reheating.
Since the decay width of the inflaton tends to be smaller than the axion mass $\Gamma_\phi < m_a$, the axion starts to oscillate during the matter dominated era driven by the inflaton $\phi$ before reheating.
In this case, the Universe becomes axion dominated when the temperature of the radiation bath falls below
\begin{align}
\label{eq-TMD}
    T_{\MD} = \frac{f_a^2 \theta_i^2}{3M_p^2} T_R 
      \sim 580~\GeV \times 
       \left(\frac{f_a\theta_i}{10^{16}~\GeV}\right)^2 
       \left(\frac{T_R}{10^{8}~\GeV}\right). 
\end{align}

Although the axion dominates, its quantum fluctuations are sub-dominant compared to those of the inflaton.
The power spectrum of the axion is estimated as~\cite{Lyth:2002my} 
\begin{align}
    \Pcal_{\zeta_a} 
     = \frac{4r_d^2}{(4+3r_d)^2}\left(\frac{H_*}{2\pi f_a \theta_i}\right)^2 
     \sim&\ \frac{m_0^2}{432\pi^4 f_a^2 \theta_i^2}
           \left(\frac{\mu}{M_p}\right)^2  
           \\ \notag 
      \sim&\  2.4\times 10^{-11} \times  
        \left(\frac{m_0}{10^{14}~\GeV}\right)^2  
        \left(\frac{10^{16}~\GeV}{f_a\theta_i}\right)^2 
        \left(\frac{\mu/M_p}{0.1}\right)^2,
\end{align}
where $H_*$ denotes the Hubble scale at horizon exit of the inflaton fluctuation\footnote{
Strictly speaking, the axion decay constant during inflation, \(f_a^{\rm inf}\), differs from its vacuum value \(f_a\) by a factor of a few due to the shift of \(\ita\), i.e., \(f_a > f_a^{\rm inf}\), see Table~\ref{tab:BPs}. However, we neglect this difference in the analysis of axion physics, as it does not significantly affect the results and can be effectively absorbed into the uncertainty of $\theta_i$.
} and
\begin{align}
    r_d := \left.\frac{\rho_a}{\rho_R}\right|_{H=\Gamma_a}
         = \left(\frac{T_\MD}{T_d} \right)^\frac{4}{3} \gg 1.   
\end{align}
Here, $\rho_a$ and $\rho_R$ denote the energy densities of the axion and radiation components, respectively.
Using Eq.~\eqref{eq:PWR_spectrum} together with the Friedmann equation in the slow-roll approximation,
\begin{align}
\frac{\Pcal_{\zeta_a}}{\Pcal_{\mathcal{R}}} 
 = \frac{8r_d^2}{(4+3r_d)^2} \left(\frac{M_p}{\theta_i f_a}\right)^2 
    \eps_V 
 \sim 0.05\times \left(\frac{10^{16}~\GeV}{\theta_i f_a}\right)^2
                 \left(\frac{\eps_V}{10^{-6}}\right),
\end{align}
where $\Pcal_{\mathcal{R}}$ is the power spectrum of the inflaton. 
Hence, the power spectrum is predominantly generated by the inflaton, but it can receive $\mathcal{O}(1)\%$ corrections depending on the parameters.

The non-Gaussianity induced by the axion is estimated as~\cite{Sasaki:2006kq} 
\begin{align}
  f_{\mathrm{NL}}^\eff 
    =\left(\frac{\Pcal_{\zeta_a}}{\Pcal_\mathcal{R}}\right)^2 
     \times \frac{5}{12}\left(-3 + \frac{4}{r_d} 
      + \frac{8}{4+3r_d}\right) 
    \sim -1.25\times10^{-2} \times 
          \left(\frac{\Pcal_{\zeta_a}/\Pcal_{\mathcal{R}}}
                      {0.1}\right)^2.
\end{align}
Since the current limit is $\order{1}$, it is negligibly small. 
In the calculation here, we assume the quadratic potential, 
but the result does not change significantly 
unless the axion resides precisely near the hill-top of the cosine potential~\cite{Kawasaki:2011pd}.

In our scenario, axion domination can last for a long time, and axions eventually decay into radiation.
The decay leads to a large entropy dilution factor, 
\begin{align}
    \Delta := \frac{s_{\mathrm{after}}}{s_{\mathrm{before}}}  
            = \frac{T_\MD}{T_d} \sim 
              10^5 \times 
              \left(\frac{T_\MD}{10^3~\GeV}\right)
              \left(\frac{10~\MeV}{T_d}\right).
\end{align}
Here, $s_{\mathrm{before}}$ ($s_{\mathrm{after}}$) is the entropy density of the thermal bath before (after) axion decay.

The non-adiabatic fluctuations of the axion are transferred to the radiation bath produced by axion decay and inherited as dark matter isocurvature perturbations when dark matter is produced from this bath.
Its size is expected to be $\Pcal_{\zeta_a}/\Pcal_{\mathcal{R}} \sim \order{1}\%$, multiplied by a conversion factor from axion to dark matter, which depends on the dark matter production mechanism. In contrast, the current observational bound is $\beta_{\mathrm{iso}} \simeq \Pcal_{\zeta_a}/\Pcal_{\mathcal{R}} < 0.038$~\cite{Planck:2018jri}. A similar discussion applies to the baryon asymmetry. Therefore, the isocurvature constraint would be important in assessing our scenario in conjunction with dark matter production and baryogenesis. We leave this for future work.

The quantities related to inflaton decay and axion physics at the benchmark points are summarized in Table~\ref{tab:BPs}. 
For illustration, we set $m_f = m_{\tilde{f}} = 10^{14}~\GeV$, $\theta_i = 0.1$ and $h=1$. The inflaton decay width is evaluated with $c_G = 1$, while that of the axion is evaluated with $\sum_G N_G = 12$, $\alpha_G=1/24$, and $k_G = 12$, where the latter two are assumed to be independent of the gauge group. We choose the overall scale of the axion potential such that the inflationary trajectory is not affected and the axion decays before BBN. With these parameters, the power spectrum sourced by the axion is at most $\order{1}\%$, and the non-Gaussianity is $\order{10^{-4}}$. 
The dilution factor exceeds $\order{10^6}$ for BP2 and BP4, whereas it is of order $\order{10^2}$ for BP1 and BP3.

If the dilution factor is $\order{10^6}$, dark matter and the baryon asymmetry may be produced from or after axion decay. One may also consider the case 
in which they are produced in excess compared to the current abundance before axion decay and subsequently diluted. 
However, this possibility may lead to excessively large isocurvature perturbations. 
Thus, the most plausible scenario would be that dark matter and the baryon asymmetry are produced from axion decay. On the other hand, such a scenario might also be allowed for the smaller dilution factor $\order{10^2}$ as realized at BP1 and BP3. 
A detailed analysis of this possibility is left for future work.

\section{Conclusions}
\label{sec:conclusion}

In this work, we study an inflationary scenario driven by moduli of a finite modular symmetry. 
The modulus potential is generated at one loop via the Coleman-Weinberg mechanism. 
The modulus $\tau$ couples to heavy VLQs through modular forms in the superpotential, such that the CW potential develops a minimum at $\mathrm{Im}\,\tau \gg 1$. At this minimum, the residual $\mathbb{Z}^T_N$ subgroup of the finite modular symmetry $\Gamma_N$ remains approximately unbroken. This Abelian discrete symmetry can be applied to explain the flavor hierarchies of quarks and leptons in the SM.

We identify a direction of the modulus, $\ita$, as the inflaton $\phi$.
In the large field region, the potential is schematically given by a double-exponential form $V \sim 1 - e^{-e^\phi}$, which leads to a sufficiently flat potential suitable for inflation. 
The flatness originates from the potential induced by a modular form with non-zero $\mathbb{Z}^T_N$ charge, 
$Y_{1_t}(\tau) \sim q^{t/N}$. 
We showed that this inflationary scenario can explain the current Planck data with a spectral index of $n_s \sim 0.965$. This prediction can be compared with the somewhat higher values of $n_s$ suggested by recent ACT analyzes, and future observations will therefore provide an important test of the scenario. Owing to the flatness of the potential, the tensor-to-scalar ratio is predicted to be $\order{10^{-5}}$, far below the current experimental sensitivity.

The inflationary observables are consistent with current observations when the minimum of the potential is located at $\ita_0 \sim 2$ in the $A_4$ modular symmetric model. This value of the modulus yields the hierarchical parameter
$\abs{q}^{1/3}=e^{-2\pi \ita_0/3}\sim \order{10^{-2}\text{--}10^{-4}}$,
which is suitable for explaining the flavor hierarchies of quarks and leptons. Thus, our result suggests the intriguing possibility that the inflaton also plays the role of a flavon. The explicit connection between cosmology and flavor structure is left for future investigation.

After inflation, the inflaton $\phi$ decays into SM gauge bosons and reheats the Universe. 
The axion $a \sim \rta$, the other direction of the modulus $\tau$, 
plays an important role in the subsequent cosmological evolution.
Under the constraints from inflationary observables, the axion mass is heavier than \(\order{\mathrm{PeV}}\), while its decay constant is \(\order{10^{16}}\,\GeV\).
Although such a heavy axion decays before BBN, its oscillation energy density can temporarily dominate the energy density of the Universe before its decay. Nevertheless, the axion fluctuations remain too small to generate the observed curvature perturbations. Hence, the axion does not act as a curvaton. The resulting long-lived axion-dominated era suggests that dark matter and the baryon asymmetry should be generated at or after axion decay. 
A detailed study of these subjects, including possible isocurvature perturbations, is left for future work.

A string-theoretic embedding of our scenario deserves further study. In particular, the modulus $\tau$ may correspond to one of the moduli that remain light after flux compactification,\footnote{Some moduli can be light in certain flux backgrounds \cite{Abe:2006xi}.} since the stabilization of many moduli is often constrained by tadpole cancellation conditions~\cite{Bena:2020xrh}. A complete construction requires a consistent stabilization of the remaining moduli, as well as a better understanding of the moduli dependence of the de Sitter uplift potential. It would be particularly interesting to realize a framework in which the inflaton modulus remains light enough to drive inflation while the other moduli are stabilized at sufficiently high scales. We leave these issues for future investigation.

\section*{Acknowledgments}
\noindent
The authors thank Yusuke Yamada for his valuable discussions.
The work of Y.A. is supported by the Center of Innovations for Sustainable Quantum AI (JST Grant Number JPMJPF2221).
The work of K.G. is supported by the Yoshida Scholarship Foundation and the Keio University Doctorate Student Grant-in-Aid Program from Ushioda Memorial Fund (Graduate school recommendation).
This work is supported in part JSPS KAKENHI Grant Numbers 
No. JP22K03601 (T.H.), 25K00222 (J.K.), and JP23K03375 (T.K.)

\appendix

\section{Explicit forms of the derivatives}
\label{app:derivatives}

The derivatives of the normalized potential $F$ 
with respect to $\phi$ are given by 
\begin{align}
 F_\phi =&\ \sqrt{\frac{2}{h}} y (k-px)  
            \left[ \log y - 1 + \left( 1+\frac{y}{z} \right)  
                            \log\left( 1+\frac{z}{y} \right) 
            \right],
\\ \notag 
 F_{\phi\phi} =&\ 
        \frac{2}{h} y 
   \left[ 
       \left\{ (k-px)^2 -px \right\}
       \left\{ \log y -1 + \left(1+\frac{y}{z}\right) 
                       \log\left(1+\frac{z}{y}\right)
       \right\} 
   \right. 
\\ \notag 
&\  \left.  \hspace{8.0cm} + 
       \left(k-px\right)^2 \frac{y}{z}\log\left(1+\frac{z}{y}\right)
  \right], 
\\ \notag 
F_{\phi\phi\phi} =&\ 
  \sqrt{\frac{8}{h^3}} y \left[ 
       -\left(k-px\right)^3 \left(1+\frac{z}{y}\right)^{-1} 
\right. \\ \notag 
&\  \hspace{0.0cm} + 
        \left\{ (k-px)^3-3px (k-px) -px \right\}  
        \left\{ 
            \log y -1 +\left(1+\frac{y}{z}\right)\log\left(1+\frac{z}{y}\right)
        \right\} 
\\ \notag 
&\ \left. \hspace{0.0cm}  +
      3(k-px)\left\{(k-px)^2 -px\right\} \frac{y}{z}
     \log \left(1+\frac{z}{y}\right) 
 \right].  
\end{align}  
Note that $x$ and $y$ are functions of $\phi$.

\section{The abundance of axions produced via inflaton decay}
\label{app:axion-abundance}

The abundance of axions produced via inflaton decay depends on the inflaton branching fraction and is typically subdominant to that generated by axion oscillations through the misalignment mechanism in the relevant parameter space. To see this explicitly, we compute the axion abundances.

The inflaton predominantly decays into an axion pair.
The abundance of axions produced via inflaton decay at reheating is given by
\begin{align}
\frac{n_{a}^{\rm decay}}{s}(T_R)
\simeq
2\frac{\rho_{\phi}}{m_\phi}\frac{1}{s(T_R)}B_a
\simeq
\frac{3T_R}{2m_\phi}B_a,
\label{eq:axion-ab1}
\end{align}
where $n_{a}^{\rm decay}$ denotes the number density of axions produced via inflaton decay, $\rho_\phi$ is the inflaton energy density at the time of its decay, and hence $\rho_\phi/m_\phi$ corresponds to the inflaton number density. \(s(T_R)=\frac{2\pi^2}{45}g_*(T_R)T_R^3\) is the entropy density of the thermal bath at reheating. In the above expression, instantaneous reheating has been assumed, so that $\rho_\phi \simeq \rho_{R}(T_R) \simeq \frac{3}{4}T_R s(T_R)$. The reheating temperature due to inflaton decay is given by $T_R \simeq (1-B_a)^{1/4}\sqrt{\Gamma_\phi M_p}$ in Eq.~\eqref{eq:TR}.
On the other hand, there is another axion mode that starts to oscillate with an initial misalignment angle $\theta_i$ at $H=m_a$ during the inflaton oscillation dominated era, whereas the inflaton decays later at $H=\Gamma_\phi < m_a$.
The axion abundance from the misalignment mechanism at reheating is then given by
\begin{align}
\frac{n_{a}^{\rm osc}}{s}(T_R) \simeq
\frac{\rho_{a}^{\rm osc}}{m_a s}(T_R) \simeq
\frac{\Gamma_\phi^2 f_a^2 \theta_i^2}{m_a s(T_R)}.
\label{eq:axion-ab2}
\end{align}
Here, $n_{a}^{\rm osc}$ denotes the number density of axions produced by coherent oscillations, and $\rho_{a}^{\rm osc} \simeq m_a n_{a}^{\rm osc}$ is the corresponding energy density. In the above calculation, we have used the evolution of non-relativistic matter in the expanding Universe, $\rho_a^{\rm osc}(T_R) = \rho_a^{\rm osc}|_{H=m_a}(\Gamma_\phi/m_a)^2$, where $\rho_a^{\rm osc}|_{H=m_a} \simeq m_a^2 f_a^2 \theta_i^2$.

Note that both abundances in Eqs.~\eqref{eq:axion-ab1} and \eqref{eq:axion-ab2} remain unchanged in the absence of entropy production, regardless of whether the axions are relativistic or non-relativistic. Therefore, the two contributions can be directly compared using ${n_a}/{s}$. More explicitly, the axions produced via inflaton decay are relativistic at reheating and eventually become non-relativistic as the Universe expands. Once they become non-relativistic, the corresponding energy abundance is given by $\rho_a^{\rm decay}/s \simeq m_a (n_a^{\rm decay}/s)$, as in the coherent oscillation case. Comparing the two contributions in Eqs.~\eqref{eq:axion-ab1} and \eqref{eq:axion-ab2}, we find that the axion abundance from coherent oscillations dominates over that from inflaton decay for
\begin{align}
\theta_i &\gtrsim \frac{M_p}{f_a} \cdot \sqrt{6 B_a(1-B_a) \frac{m_a}{m_\phi}}  
\nn \\
&\simeq 5.9\times10^{-3} \times \bigg(\frac{10^{17}~{\rm GeV}}{f_a}\bigg) 
\bigg(\frac{m_a}{10^6~{\rm GeV}}\bigg)^{1/2}
\bigg(\frac{10^{13}~{\rm GeV}}{m_\phi}\bigg)^{1/2}
\bigg(\frac{B_a(1-B_a)}{0.1}\bigg)^{1/2}. 
\end{align}
For $\theta_i \gtrsim \mathcal{O}(0.1)$, it is sufficient to consider only axion oscillations.

{\small
\newcommand{\arxivfont}{\rmfamily}
\bibliographystyle{yautphysm}
\bibliography{refs}
}

\end{document}